\documentclass[
preprint,
superscriptaddress,
longbibliography
]{revtex4-2}

\usepackage{graphicx}
\usepackage{dcolumn}
\usepackage{bm}
\usepackage{epstopdf}
\usepackage{xr}
\usepackage{amsmath}
\usepackage{color}
\usepackage{comment}
\usepackage[normalem]{ulem}
\usepackage{wasysym}
\usepackage{siunitx}
\usepackage{physics}
\usepackage{gensymb}
\usepackage{lineno}
\usepackage[T1]{fontenc}
\usepackage{booktabs}
\usepackage[section]{placeins}
\usepackage{textcomp}
\usepackage{amssymb}
\usepackage{gensymb}
\usepackage{xcolor}
\DeclareUnicodeCharacter{0306}{ }
\usepackage{textcomp}
\usepackage{textcomp}
\usepackage{upgreek}
\usepackage[explicit]{titlesec}
\usepackage{amsmath, upgreek}

\usepackage{hyperref}
\hypersetup{colorlinks=true,%
citecolor=blue,%
filecolor=black,%
linkcolor=black,%
urlcolor=blue
}

\renewcommand{\figurename}{\textbf{Fig.}}

\makeatletter

\newcommand{\Rmnum}[1]{\expandafter\@slowromancap\romannumeral #1@}
\makeatother

\begin{document}

\title{Quantum interference in a twisted high-Tc SQUID senses emergent interfacial order}
    \author{Amit Basu}
    \homepage{amitbasu457@gmail.com}
    \affiliation{Department of Condensed Matter Physics and Materials Science, Tata Institute of Fundamental Research, Homi Bhabha Road, Mumbai 400005, India}
    \author{Samrat Ash}
    \affiliation{Department of Condensed Matter Physics and Materials Science, Tata Institute of Fundamental Research, Homi Bhabha Road, Mumbai 400005, India}
     \author{Ritajit Kundu}
    \affiliation{Department of Physics, Indian Institute of Technology Kanpur, Kanpur 208016, India.}
    \author{Neha Bhatia}
    \affiliation{Department of Condensed Matter Physics and Materials Science, Tata Institute of Fundamental Research, Homi Bhabha Road, Mumbai 400005, India}
    \author{Sakshi Nema}
    \affiliation{Department of Condensed Matter Physics and Materials Science, Tata Institute of Fundamental Research, Homi Bhabha Road, Mumbai 400005, India}
    \author{Tejaswini Gawade}
    \affiliation{Department of Condensed Matter Physics and Materials Science, Tata Institute of Fundamental Research, Homi Bhabha Road, Mumbai 400005, India}
    \author{Khushabu Agrawal}
    \affiliation{Department of Condensed Matter Physics and Materials Science, Tata Institute of Fundamental Research, Homi Bhabha Road, Mumbai 400005, India}
    \author{Abhishek Das}
    \affiliation{Department of Condensed Matter Physics and Materials Science, Tata Institute of Fundamental Research, Homi Bhabha Road, Mumbai 400005, India}
    \author{Joydip Sarkar}
    \affiliation{Department of Condensed Matter Physics and Materials Science, Tata Institute of Fundamental Research, Homi Bhabha Road, Mumbai 400005, India}
    \author{Amit Shah}
    \affiliation{Department of Condensed Matter Physics and Materials Science, Tata Institute of Fundamental Research, Homi Bhabha Road, Mumbai 400005, India}
    \author{Ruta Kulkarni}
    \affiliation{Department of Condensed Matter Physics and Materials Science, Tata Institute of Fundamental Research, Homi Bhabha Road, Mumbai 400005, India}
    \author{Digambar A. Jangade}
    \affiliation{Department of Condensed Matter Physics and Materials Science, Tata Institute of Fundamental Research, Homi Bhabha Road, Mumbai 400005, India}
    \author{Arijit Kundu}
    \affiliation{Department of Physics, Indian Institute of Technology Kanpur, Kanpur 208016, India.}
    \author{A. Thamizhavel}
    \affiliation{Department of Condensed Matter Physics and Materials Science, Tata Institute of Fundamental Research, Homi Bhabha Road, Mumbai 400005, India}   
    \author{Mandar M. Deshmukh}
    \homepage{deshmukh@tifr.res.in}
    \affiliation{Department of Condensed Matter Physics and Materials Science, Tata Institute of Fundamental Research, Homi Bhabha Road, Mumbai 400005, India}

\maketitle

\newpage

\textbf{Engineering artificial systems by twisting and stacking van der Waals materials has proven to be an excellent platform for exploring emergent quantum phenomena that can be significantly different from the constituents. Recent advances in the fabrication of high-quality twisted interfaces provide a unique opportunity to study the little-explored interfacial superconducting order in twisted cuprate superconductors. In our work, we fabricate superconducting quantum interference devices (SQUID) that utilize the twisted interface of $\mathbf{Bi_2Sr_2CaCu_2O_{8+\delta}}$, a high-Tc cuprate superconductor. By measuring the magnetic field modulation of switching current and differential resistance, we find a $\mathbf{\pi}$ phase difference between the two Josephson junction arms of the SQUID reflecting chiral superconducting order -- a crucial aspect inaccessible to single Josephson junction devices of the past. Our observations also indicate co-tunneling of the Cooper pairs and a time-reversal symmetry-broken emergent superconducting order. Additionally, these SQUIDs are well suited for use as state-of-the-art flux sensors close to 77 K, achieving a flux noise sensitivity of $\sim$1.5 \textbf{$\upmu \Phi_0$/$\sqrt{\mathrm{Hz}}$}. Stabilizing new superconducting orders using twisted interfaces and probing them using quantum interference opens new avenues to understanding the microscopic origin of unconventional superconductors. Our SQUID architecture is suitable for investigating the charge transport mechanisms and the symmetry of superconducting order at the interfaces of other systems, reflecting the broad applicability beyond cuprate superconductors.}

\vspace{5mm}

Macroscopic phase coherence in a superconductor is the central property that results in observable properties like the dissipationless flow of Cooper pairs and the perfect diamagnetic response. The crucial role of the macroscopic phase coherence of Cooper pairs in a superconductor is observed in a Josephson junction (JJ) and in a pair of Josephson junctions in a parallel configuration; the latter device is the superconducting quantum interference device (SQUID). The quantum interference in the SQUID is tuned by an external magnetic field threading the loop formed by the JJs. JJ and SQUID motifs are central for realizing superconducting qubits, quantum sensors, and amplifiers 
\cite{clarke_superconducting_2008,devoret_superconducting_2013,fagaly_superconducting_2006,SQUIDhandbook1,vasyukov_scanning_2013}. Magnetic flux and field sensors using the SQUID play a key role in many applications as interference is modulated by a single quantum of flux that threads the SQUID loop \cite{SQUIDhandbook1,SQUIDhandbook2}. In addition to phase coherence, the symmetry of the superconducting order parameter plays a crucial role in realizing SQUID response. Most superconducting electronics use s-wave superconductors, where the order parameter is isotropic in momentum space, and the critical temperature ($T_c$) is low. Whereas the high-$T_c$ superconductors (HTSC) have an anisotropic order parameter with a \textit{d}-wave symmetry and a high transition temperature. The anisotropy of the superconducting order parameter in momentum space offers the opportunity to use JJs incorporating the relative angular orientation of superconductors to shed light on the interfacial order. By introducing a twist angle, new types of devices can be developed for technology \cite{can_high-temperature_2021, patel_d_2024,brosco_superconducting_2024,coppo_flux-tunable_2024} and to understand the microscopic origins of HTSC.

Recently, there has been a growing interest in studying heterostructures of various two-dimensional (2D) van der Waals (vdW) materials by introducing a relative twist angle between the layers. The advantage of vdW materials is that they can be easily exfoliated down to a few unit cell thickness limit due to the weak interlayer vdW interaction; this opens up the possibility of exploring a plethora of novel artificial quantum devices proposed theoretically \cite{can_high-temperature_2021} and realized experimentally \cite{zhao_time-reversal_2023, ghosh_high-temperature_2024,narita_field-free_2022,ando_observation_2020,wu_field-free_2022,farrar_superconducting_2021,portoles_tunable_2022}. The twist angle between the stacked layers plays a crucial role in such systems, as it determines the hybridization between the two layers and electronic interactions. Examples of these twisted systems are those based on graphene systems \cite{cao_unconventional_2018}, twisted transition metal dichalcogenides (TMDC) \cite{guo_superconductivity_2025,xia_superconductivity_2025}, and cuprate superconductors \cite{zhao_time-reversal_2023,lee_encapsulating_2023,ghosh_high-temperature_2024,Qi-kunPhysRevX.11.031011}. In many of these heterostructures, the band structure is tuned with a twist angle, and resulting flat bands host correlated electronic states and superconductivity -- an aspect that is widely studied. In contrast, studies on twisted cuprate high-$T_c$ systems are in their initial stages \cite{Qi-kunPhysRevX.11.031011,zhao_time-reversal_2023,lee_encapsulating_2023, ghosh_high-temperature_2024,martini_twisted_2023} and show promise to observe emerging interfacial superconducting orders with implications for fundamental science \cite{pixley_twisted_2025, confalone_cuprate_2025}.

\begin{figure*}[h]
\includegraphics[width=15.5cm]{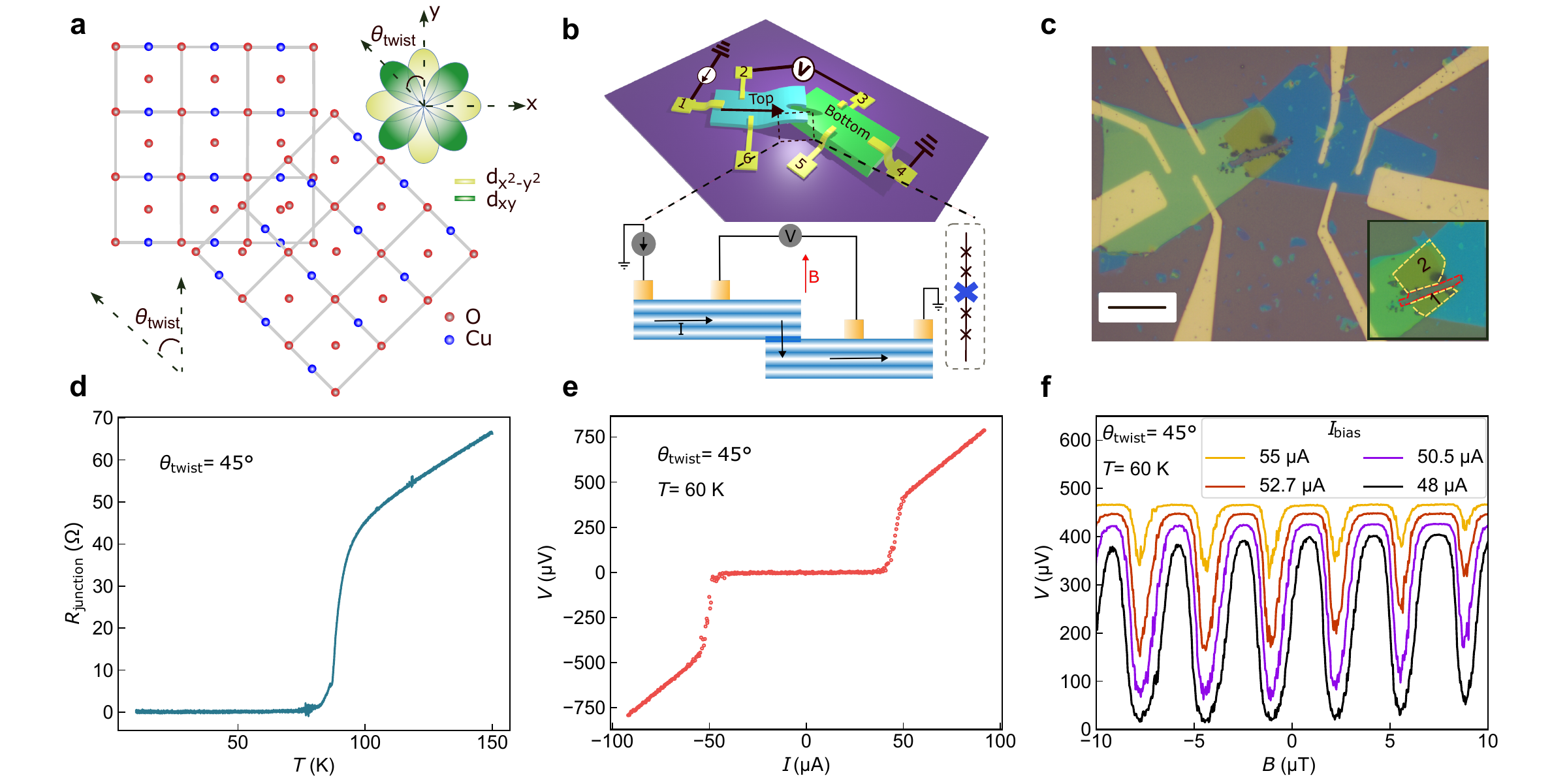}
\caption{ \label{fig:fig1} \textbf{Device Schematic and basic electrical characterization of SQUID.} (a) Schematic of two Cu$\mathrm{O}$ layers, twisted by an angle $\theta_{\mathrm{twist}}$ with respect to each other. The inset shows the schematic of the order parameter close to the twisted interface. (b) Schematic of a twisted BSCCO SQUID. The current is sourced from electrode \textbf{1} and grounded at electrode \textbf{4}; the voltage drop across the junction is measured between electrodes \textbf{2} and \textbf{3}. The inset shows the current flow across the junction; c-axis current flow comprises the flow through intrinsic Josephson junctions and artificial Josephson junction. The artificial Josephson junction is highlighted in dark blue. (c) Optical micrograph of a 45\textdegree~twisted SQUID. The scale bar is 20 $\upmu$m. In the inset, the zoomed-in SQUID loop and parallel JJ arms are shown. SQUID loop is marked with a red dashed line, whereas the outline of the two JJ arms is marked with a yellow dashed line. (d) Temperature dependence of the four-probe resistance across the junction. Resistance is measured using an $ac$ current of 100 nA. (e) The $dc$ $I-V$ characteristics of the 45\textdegree~twisted SQUID at 60 K. (f) Modulation of the voltage drop across the junction with perpendicular magnetic field, measured at different fixed $dc$ bias currents for a 45\textdegree~twisted device at 60 K.}

\end{figure*}

Bi$_2$Sr$_2$CaCu$_2$O$_{8+\delta}$ (Bi-2212) or BSCCO is a well-known high-$T_c$ ($T_c \sim 92$~K) vdW cuprate superconductor. The heart of the HTSC of BSCCO is CuO planes \cite{yu_high-temperature_2019,sterpetti_comprehensive_2017} that have a \textit{d}-wave order parameter, as shown in ~\hyperref[fig:fig1]{Fig. 1(a)}. It is exfoliable down to a single unit cell and is amenable to making twisted heterostructures. A major breakthrough in the experimental study of twisted BSCCO was the development of the cryogenic exfoliation technique by Zhao \textit{et al.} \cite{zhao_time-reversal_2023}; this process preserves the oxygen stoichiometry during the exfoliation and assembly of BSCCO interfaces. Following this, recent studies of twist angle-dependent transport have shown remarkable properties like the signature of time-reversal symmetry broken superconducting order \cite{zhao_time-reversal_2023} and field-tunable Josephson diode effect \cite{ghosh_high-temperature_2024}. For a 45° twisted interface, Cooper-pair tunneling is suppressed, and the supercurrent becomes vanishingly small \cite{zhao_time-reversal_2023,ghosh_high-temperature_2024}. As a result, the critical current density of a 45° twisted device reduces significantly compared to the untwisted interfacial junction — consistent with theoretical predictions \cite{can_high-temperature_2021}. However, co-tunneling of Cooper pairs (tunneling of two Cooper pairs) is allowed by order parameter symmetry and is proposed as a possible charge-transport mechanism \cite{can_high-temperature_2021,song_doping_2022,volkov_josephson_2024}. The dominant role of co-tunneling can lead to two minima in the free energy landscape and result in modification of the current phase relationship (CPR). Indirect signatures of the modified CPR with second harmonic have been observed in a twisted JJ formed using BSCCO flakes \cite{zhao_time-reversal_2023}. The CPR \cite{endres_currentphase_2023,babich_limitations_2023,messelot_direct_2024,murani_ballistic_2017} and the anomalous phase difference \cite{wollman_experimental_1993} can be directly probed only using the quantum interference observed in SQUIDs. We exploit the quantum interference in SQUID geometry to sense the anomalous phase difference when the two JJs host distinct chiral superconducting orders and to reveal the harmonic content in the CPR of a JJ. Additionally, complete control of the twist angle is realized in our devices, in contrast to past studies of high-Tc grain boundary SQUIDs \cite{koelle-RevModPhys.71.631,faley_high-_2017,schneider_half-_2004,hilgenkamp_grain_2002,dantsker_addendum_1995}.   

In this work, we have studied the twisted interface of BSCCO for different twist angles (0\textdegree, 40\textdegree, 42\textdegree, 44\textdegree, 45\textdegree ~devices, details of the devices are provided in supplementary information \textbf{SI 3}) by making asymmetric SQUID devices using a simple and effective technique to create a SQUID loop (see Methods and \textbf{SI 1}). We observe a clear signature of quantum interference as the magnetic field threaded through the SQUID loop is varied. Crucially, we directly observe an anomalous phase difference that results from superconducting phases of opposite chirality in the two JJs in the SQUID loop. We can also infer a prominent second-harmonic contribution in the CPR and breaking of the time-reversal symmetry. We will discuss these results in detail in the following sections. In addition to fundamental physics, we also show that our device performs with state-of-the-art specifications as a SQUID with flux noise $\sim$ 1.5 $\mathrm{\upmu\Phi_0/\sqrt{Hz}}$ at 60 K, where $\Phi_0$ denotes the magnetic flux quantum.

The primary challenge in the fabrication of a \textit{c}-axis SQUID is to create overlapping layers of BSCCO flakes with known precise twist angle. We follow the widely used strategy of starting with a mother flake and cleaving it along the \textit{c}-axis to derive two daughter flakes from it at cryogenic temperatures \cite{zhao_time-reversal_2023}. By introducing a relative twist between the daughter flakes and overlapping them produces a c-axis JJ. We have used PDMS-assisted dry cryogenic exfoliation \cite{zhao_time-reversal_2023, ghosh_high-temperature_2024,martini_twisted_2023,lee_encapsulating_2023}  technique to fabricate a pristine twisted interface for SQUID devices. We make Au contacts with a shadow SiN stencil mask \cite{Zhao_PhysRevLett.122.247001}; these electrodes are extended by depositing Cr/Au using a metal mask. We ensure that Cr does not come into contact with the BSCCO flakes \cite{ghosh_demand_2020}.  Finally, we create the SQUID geometry by making an asymmetric cut across the junction (see supplementary video 1) using a sharp fiber tip \cite{varma_sangani_facile_2020} (fabrication details are provided in the \textbf{SI 1} and Methods section). In \textbf{SI 2}, we show that cutting does not cause additional folding or any other damage to the stack. Right after the completion of fabrication, we load our device in a closed-cycle cryostat (cryostat details are in the \textbf{SI 4}). The measurements of the devices are done in a modified probe station with a cryoperm shield to protect the device from Earth's magnetic field. Inside the cryoperm-shielded box, a two-axis vector electromagnet made of copper wires is designed to apply a magnetic field (see \textbf{SI 4}, for details of the experimental setup).

Electrical measurements are done using a low-noise setup (details in the \textbf{SI 4}). We have performed $dc$ transport measurements of the SQUID in a four-probe configuration, as shown in the schematic of \hyperref[fig:fig1]{Fig. 1(b)}. Current flows through the junction along the \textit{c}-axis (shown in the inset of \hyperref[fig:fig1]{Fig. 1(b)}). In the BSCCO crystal structure, JJs are present along \textit{c}-axis, called intrinsic Josephson junction (IJJ). When we cleave one flake into two daughter flakes and make the twisted stack, an artificial Josephson junction (AJJ) is formed at the interface (the interface of the two BSCCO flakes shown in the inset of \hyperref[fig:fig1]{Fig. 1(b)}, marked dark blue). When the current flows through the junction along \textit{c}-axis, it passes through a series of IJJs and AJJ. The critical current density of AJJ is much smaller than that of IJJ \cite{ghosh_high-temperature_2024}, and probing the device at low current only probes the AJJ formed at the twisted interface. 

Here, we will discuss the SQUID oscillation observed for two different devices with a twist angle ($\theta_{\mathrm{twist}}$) 45\textdegree ~ and 40\textdegree~(additional devices are documented in \textbf{SI 9} and \textbf{SI 14}). The optical image of 45\textdegree~twisted BSCCO SQUID device is shown in \hyperref[fig:fig1]{Fig. 1(c)}. SQUID loop and the two constituting JJ arms of the SQUID are shown in the inset of \hyperref[fig:fig1]{Fig. 1(c)}. The area and perimeter of the loop are crucial for understanding the SQUID response of the twisted interface, an aspect that we discuss later. Initially, we characterize the junction quality based on two parameters: critical temperature and critical current density \cite{pixley_twisted_2025}; these numbers are crucial to understand the SQUID response. The temperature dependence of four-probe resistance across the junction is measured using an $ac$ current ($\sim$100~nA) through the 45\textdegree~SQUID, shown in \hyperref[fig:fig1]{Fig. 1(d)}. The critical temperature of the junction is $\sim$75~K. We measure the $dc$ $I-V$ characteristics of the junction to confirm the quality of AJJs, as shown in \hyperref[fig:fig1]{Fig. 1(e)}. The critical current density of the 45\textdegree~twisted SQUID is 0.013 kA/$\mathrm{cm^2}$ at 60 K. This value is comparable with previously reported twisted BSCCO JJ studies \cite{zhao_time-reversal_2023, ghosh_high-temperature_2024} (see comparison in \textbf{Table III}, \textbf{SI 6}) and significantly lower relative to that of the untwisted junctions (comparison table \textbf{Table IV} in \textbf{SI 6}). It reflects the suppression in Cooper-pair tunneling due to the relative rotation of order parameters at the interface of the two flakes.

\begin{figure*}[h]
\includegraphics[width=15.5cm]{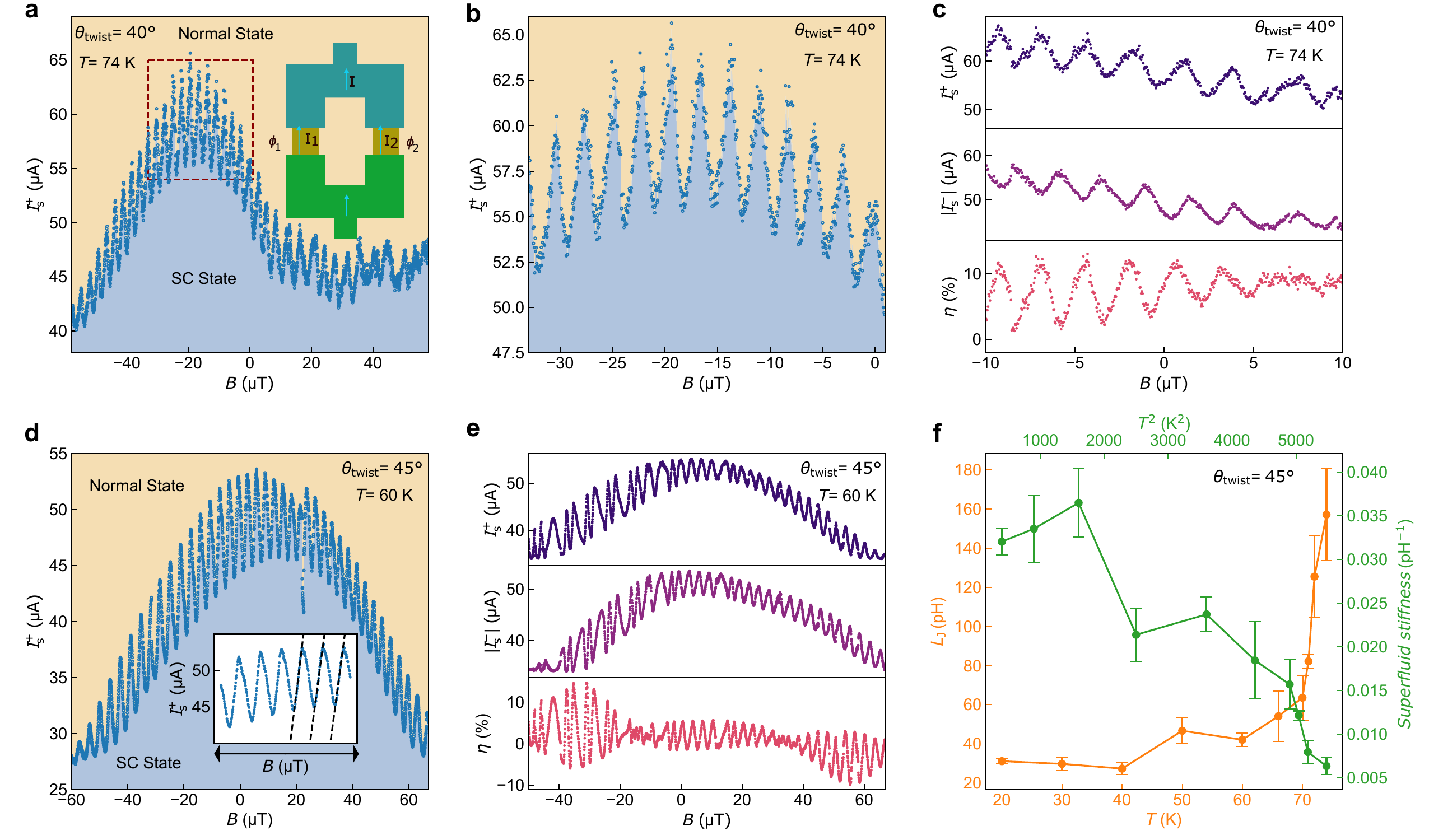}
\caption{ \label{fig:fig2} \textbf{Superconducting quantum interference in a twisted BSCCO SQUID.} (a) Field dependence of positive side switching current is measured by applying a magnetic field perpendicular to the flake at 74 K for a 40\textdegree~device. SQUID loop area is 131 $\upmu \mathrm{m^2}$, and its perimeter is 54 $\upmu \mathrm{m}$. Expected field modulation period is $\sim 15$ $\upmu \mathrm{T}$.   SQUID oscillations are present within a smooth envelope. The blue-shaded region defines the superconducting state (SC state) of the junction. In the inset, the schematic shows the phase drop across two JJ arms of the SQUID, bias current flowing across each arm, and the external flux threading the SQUID loop. (b) Enlarged view of the region highlighted by the dark red dashed box in (a). The perpendicular field response of the switching current of the large AJJ arm forms the outer envelope. (c) Positive ($I_s^+$) and negative (|$I_s^-$|) switching current measurement with out-of-plane magnetic field and resulting diode asymmetry ($\eta$) is demonstrated at 74 K for 40\textdegree~twisted device. (d) Modulation of switching current with the perpendicular field is shown at 60 K for a 45\textdegree~twisted device. SQUID loop area is 81 $\upmu \mathrm{m^2}$, and its perimeter is 56 $\upmu \mathrm{m}$. Expected field modulation period is $\sim 25$ $\upmu \mathrm{T}$. In the inset, slope extraction from the switching current modulation is shown, which allows to calculate the Josephson inductance. (e) For a 45\textdegree~twisted device, positive and negative switching current modulation with applied magnetic field and resulting diode asymmetry at 60 K is demonstrated. (f) The temperature dependence of the Josephson inductance ($L_J$) is shown with an error bar. In the same plot, superfluid stiffness is plotted as a function of $T^2$.} 
\end{figure*}

One of the hallmarks of a SQUID is the quantum interference between the two arms of the device, analogous to the double-slit experiment in optics. The interference of the macroscopic quantum phase manifests in two distinct ways: (i) modulation of the voltage drop across the SQUID at a fixed bias current as a function of magnetic field, and (ii) modulation of the switching current ($I_s$) from the dissipationless to the dissipative state, also as a function of magnetic field. \hyperref[fig:fig1]{Fig. 1(f)} shows the modulation of the voltage drop for different current bias points on the \textit{dc} $I-V$ curve with magnetic field. In addition, we have measured the switching current using a counter-based technique \cite{murani_ballistic_2017, ghosh_high-temperature_2024} (details of the measurement protocol are given in \textbf{SI 5}).
In \hyperref[fig:fig2]{Fig. 2(a)} and \hyperref[fig:fig2]{Fig. 2(d)}, we show the modulation in $I_s$ for a 40\textdegree~and 45\textdegree~twisted device, respectively, with an out-of-plane magnetic field. The switching current of the smaller arm is modulated due to the applied magnetic field, whereas the larger arm creates the outer envelope of the modulation. The observed magnetic field period of the switching current modulation is $2.2 ~ \mathrm{\upmu T}$ for 40\textdegree~twsited device (see \hyperref[fig:fig2]{Fig. 2(a)}) and $3.3 ~ \mathrm{\upmu T}$ for 45\textdegree~twisted device (see \hyperref[fig:fig2]{Fig. 2(d)}). We note that this period is approximately seven times smaller, for both twist angles, compared to the expected period ($B_{period}=\Phi_0/A_{geo}$, ) derived from the geometrical area $A_{geo}$ of the loop; this is surprising and unusual (see \textbf{Table V} in \textbf{SI 7} for the period data from other devices). The reduced period could be due to two possible mechanisms: first, the Meissner screening currents widen the effective area, and second, the flux-focusing effect concentrates the flux lines from the superconducting region to the loop area, increasing the effective field. Details of these mechanisms are given in \textbf{SI 7}. In \hyperref[fig:fig2]{Fig. 2(a)} inset, a schematic of a SQUID and phase drop across its two JJ arms is shown. For SQUID with negligible geometric inductance, phase drop across arm 1 (consisting of JJ1) and arm 2 (consisting of JJ2) are $\phi_1$ and $\phi_2$ respectively, constrained by the applied flux $\Phi_{\mathrm{ext}}$ as $\phi_1-\phi_2=2\pi \dfrac{\Phi_{\mathrm{ext}}}{\Phi_0}$. As we will discuss later, finite geometrical inductance will modify this phase sum rule.
\hyperref[fig:fig2]{Fig. 2(b)} shows the zoomed-in SQUID oscillations of the 40\textdegree~twisted device.

To understand the response of the SQUID, we have measured both the switching current at positive bias, $I_s^+$, and negative bias, $I_s^-$, at each magnetic field value during the field sweep. Switching data for the 40\textdegree~twisted device is presented in \hyperref[fig:fig2]{Fig. 2(c)} and for the 45\textdegree~twisted device is shown in \hyperref[fig:fig2]{Fig. 2(e)}. We define SQUID-based diode efficiency as  $\eta=\dfrac{I_s^+-|I_s^-|}{I_s^++|I_s^-|} \times100 \%$. Here, the observed diode effect is tunable with a perpendicular magnetic field and arises in the SQUID architecture for the data shown in \hyperref[fig:fig2]{Fig. 2(c)} and \hyperref[fig:fig2]{Fig. 2(e)}; diode asymmetry in SQUIDs has been theoretically \cite{souto_josephson_2022,cuozzo_microwave-tunable_2024} and experimentally \cite{li_interfering_2024} studied in the past. Notably from the RCSJ model \cite{tinkham_superconductivity}, $I_s^+$ and $I_s^-$ should be of the same magnitude. But breaking of time-reversal symmetry and inversion symmetry lead to the mismatch between $I_s^+$ and $|I_s^-|$, known as the superconducting diode effect -- this is a crucial asymmetry at $B=0$, we discuss later. The field response of other devices with different twist angles is shown in \textbf{SI 9}. The nominally 0\textdegree~twisted SQUID does not show a diode effect at zero magnetic field and serves as a crucial control. 

Another important aspect of the SQUID architecture is the inductances associated with its JJ arms. The inductances primarily have two origins: geometric inductance of the loop and Josephson inductance ($L_J$). The extended SQUID arm introduces finite geometric inductance, which modifies the SQUID response from the ideal behavior expected from the CPR. The phase sum rule for the SQUID geometry is 
\begin{equation}
\label{eq:eq1} 
    2\pi(\dfrac{\Phi_{\mathrm{ext}}-(L_{geo,1}I_{1}(\phi_1) - L_{geo,2}I_{2}(\phi_2))}{\Phi_0})= \phi_1 - \phi_2
\end{equation} 
where $\phi_1$ and $\phi_2$ are the phase drop across individual JJs, labeled as JJ1 and JJ2, and $L_{geo,1}$ and $L_{geo,2}$ are the geometric inductances of the two arms, and $I_{1}(\phi_1),$ $I_{2}(\phi_2)$ are the supercurrents flowing through SQUID arm 1 and arm 2 respectively and the total current flowing through the SQUID is $I=I_{1}+I_{2}$. We have calculated the geometric inductance using electromagnetic simulations (see \textbf{SI 8})  to quantify it; this allows us to probe the Josephson inductances and the associated physics. We note that the kinetic inductance of the pristine flake is < 1 pH, shown in \textbf{SI 13}. So the contribution of kinetic inductance is neglected. For 45\textdegree~SQUID device, shown in \hyperref[fig:fig1]{Fig. 1(c)}, the areal asymmetry between two arms of the 45\textdegree~SQUID is 3:1. We find that the two arms of the SQUID are asymmetric with finite geometric inductance (see \textbf{SI 8} for details). The phase drop across the small junction (JJ1), $\phi_1$, is tuned with the applied flux, while the phase drop across the large junction (JJ2), $\phi_2$, is nearly constant. In the counter-based measurement technique, we can measure the supercurrent at which the interfacial JJ switches from superconducting to normal state, in that case $I$ becomes $I_\mathrm{s}$ and $I_{1(2)}$ becomes $I_\mathrm{s,1(2)}$. Switching current of junction 2, $I_{s2}$ is almost constant (as $I_{s2}$>>$I_{s1}$), so the switching current of arm 1 is $I_{s1}=\dfrac{\Phi_{\mathrm{ext}}+L_{geo,2}I_{s2}+\dfrac{\Phi_0}{2\pi}\phi_2}{L_{J,1}+L_{geo,1}}$ (derived from \hyperref[eq:eq1]{Eq. 1}); where the Josephson inductance of arm 1 is $L_{J,1}=\dfrac{\Phi_0}{2\pi} \dfrac{\phi_1}{I_{s1}}$. Therefore, Josephson inductance of the smaller arm can be extracted from the slope of the flux modulation of the switching current \cite{portoles_tunable_2022, PhysRevLett.134.216001}. From the slope calculation shown in the inset of \hyperref[fig:fig2]{Fig. 2(d)} and using the relation $\dfrac{\delta I_{s,1}}{\delta \Phi_{\mathrm{ext}}}= \dfrac{1}{L_{J,1}+L_{geo,1}}$ we get the $L_{J,1}+L_{geo,1}$ of our device. We then use electromagnetic simulation to quantify (see \textbf{SI 8}) the effect of geometric inductance ($L_{geo,1}$) in the SQUID response with a goal to measure the Josephson junction inductance. We have also performed numerical simulations to extract the switching current and inductances starting with the CPR, shown in \textbf{SI 11}.

Using the analysis described above, we now extract the Josephson inductance of the twisted junction as a function of temperature, as shown in \hyperref[fig:fig2]{Fig. 2(f)}. The increase in Josephson inductance with increasing temperature is qualitatively similar to previous experiments on bulk BSCCO \cite{osborn2001superfluid}; however, importantly, we probe the inductance of the interfacial JJ and not that of the bulk. The superfluid stiffness, which is inversely proportional to the Josephson inductance, quantifies the strength of phase coherence in a superconductor. The temperature scaling of the superfluid stiffness for the interfacial JJ differs from that of the bulk BSCCO. In bulk BSCCO, the superfluid stiffness decreases linearly as a function of $T^2$ \cite{osborn2001superfluid} and is clearly different from the response we measure for the twisted interfacial junction (see green trace in \hyperref[fig:fig2]{Fig. 2(f)}). The distinct response of the JJ provides a hint of the unusual nature of the interfacial superconductor, which will form the focus of our discussion next.

Now, we discuss the key aspect of our SQUID-based experiment to probe any anomalous phase at the twisted interface. To probe any interfacial phase differences, it is desirable to have a minimum geometrical inductance. However, in our devices, the geometric inductance and the Josephson inductance are comparable. In other words, the screening parameter $\beta = 2\pi \frac{L_{geo}I_c}{\Phi_0}$ $\sim 1 $; here, $I_c$ is the critical current of the SQUID. \hyperref[eq:eq1]{Eq. 1} can be rewritten (see method section) \cite{wollman_experimental_1993}  after including an anomalous phase difference $\phi_{12}$ as
\begin{equation}
\label{eq:eq2} 
    \phi_1 -\phi_2 + \mathcal{A}~L_{geo}~I +\phi_{12}=2\pi \frac{\Phi_{\mathrm{ext}}}{\Phi_0}
\end{equation}
 where $\phi_1$ and $\phi_2$ are phase drop across the two junctions, $L_{geo}$ is the combined geometrical inductance of the loop, $I$ is the \textit{dc} current bias of the SQUID, and $\mathcal{A}$ is a constant that quantifies the asymmetry of the two JJs and the inductance of the individual arms. Crucially, $\phi_{12}$ is a new term compared to \hyperref[eq:eq1]{Eq. 1} and accounts for the intrinsic phase difference between the two JJ arms of the SQUID due to the interfacial order in each of the JJs. The advantage of \hyperref[eq:eq2]{Eq. 2} is that it allows us to extrapolate to the $I=0$ limit in order to quantify $\phi_{12}$ even though $L_{geo}$ is not negligible \cite{wollman_experimental_1993}, as is the case in our devices. In the simplest case where the two junctions are copies of each other $\phi_{12}=0$, modulo $2\pi$. However, if the two junctions are distinct in the quantum mechanical origin of the chiral order, then $\phi_{12}\neq 0$, and this can reveal information about the JJs only using the SQUID.

\begin{figure*}[h]
\includegraphics[width=15.5cm]{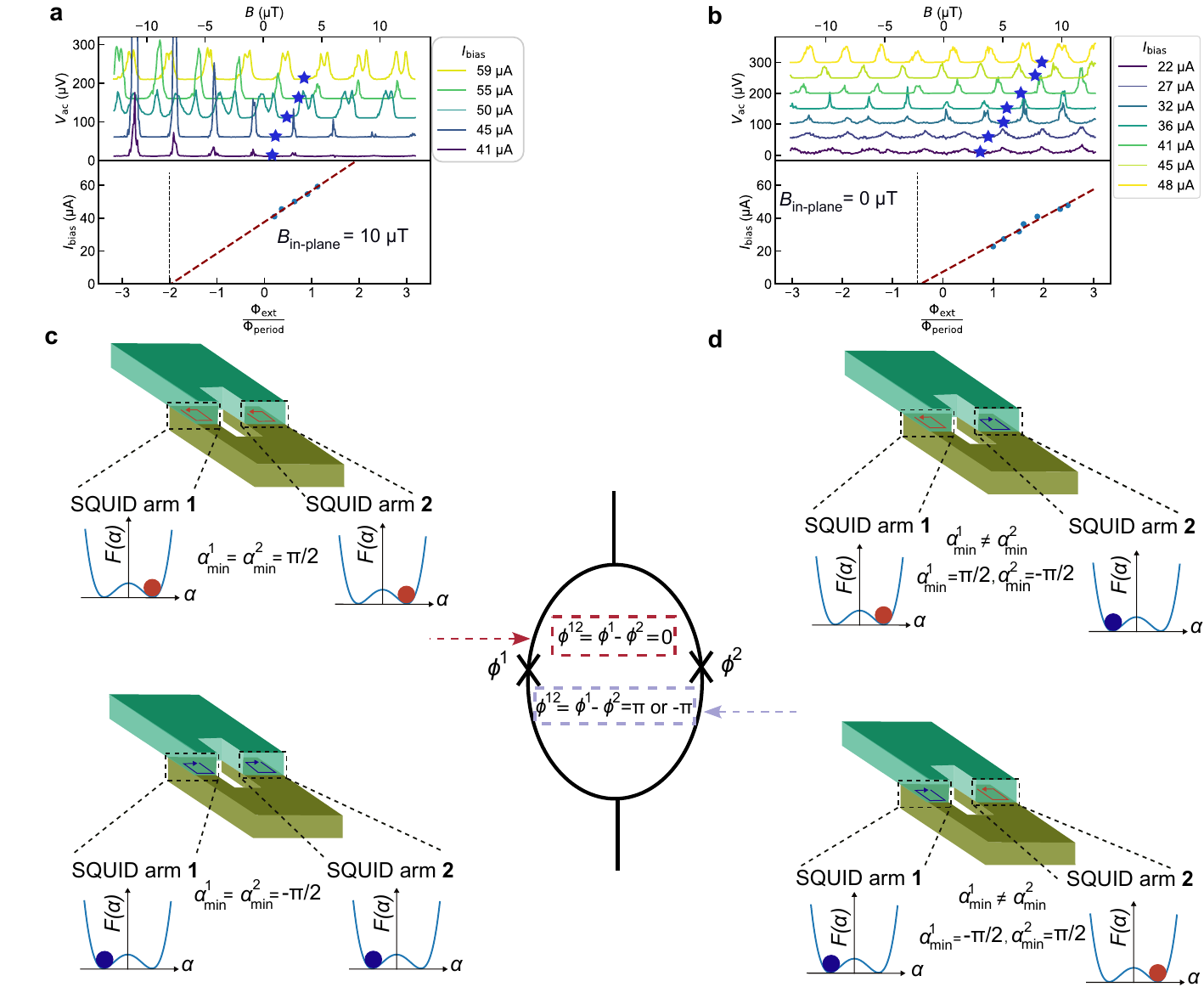}
\caption{ \label{fig:fig3} \textbf{Signature of chiral superconducting order for 45\textdegree~twisted junction.} (a) We measured \( dV/dI \) using \textit{ac} bias current (\( 1~\upmu\mathrm{A} \)) superimposed on a fixed \textit{dc} bias (\( I_\mathrm{bias} \)). The \( dV/dI \) response modulates with a perpendicular magnetic field for a fixed in-plane field value of 10 $\upmu \mathrm{T}$. The top panel shows the field dependence of $V_\mathrm{ac}$ ($\propto dV/dI$) at 20~K for various \( I_\mathrm{bias} \) (see legend), with phase minima of the modulation marked by blue stars. The bottom panel shows the plot of \( I_\mathrm{bias} \) and the corresponding phase minima (normalized by \( \Phi_\mathrm{period} \)). Here, the intercept shows a zero phase difference. (b) Similar $dV/dI$ modulation as (a), but with a fixed in-plane magnetic field value of 0 $\upmu \mathrm{T}$. Here, the intercept shows a half-period phase difference. (c) Interfacial chiral order of the same chirality (in SQUID arm \textbf{1} \& \textbf{2} are shown by red or blue), in the same free energy minima, $\alpha^1_{\mathrm{min}}= \alpha^2_{\mathrm{min}}=\pm \dfrac{\pi}{2}$. The anomalous phase difference measured by the SQUID is zero. (d) Interfacial chiral order of opposite chirality (in SQUID arm \textbf{1} \& \textbf{2} are shown by red and blue and vice versa), implying two JJs in opposite free energy minima, $\alpha^1_{\mathrm{min}}= \dfrac{\pi}{2}$, $  \alpha^2_{\mathrm{min}}=-\dfrac{\pi}{2}$ or $\alpha^1_{\mathrm{min}}= -\dfrac{\pi}{2}$, $  \alpha^2_{\mathrm{min}}=\dfrac{\pi}{2}$, resulting in anomalous phase difference $\pi$ or $-\pi$.}
\end{figure*}

We can experimentally measure the consequence of \hyperref[eq:eq2]{Eq. 2} by measuring the differential resistance ($dV/dI$) below the switching current of the SQUID following Wollman \textit{et al.} \cite{wollman_experimental_1993}. \hyperref[fig:fig3]{Fig. 3(a)}(top panel) shows several traces of the $dV/dI$ as a function of the perpendicular field for different bias currents for a 45\textdegree~twisted device. Here, as the total current through the SQUID, $I$, is varied, the phase drop across the geometric inductance of the device, $\mathcal{A}L_{geo}I$, term of \hyperref[eq:eq2]{Eq. 2} is tuned. We track the relative phase of the minima of these $dV/dI$ oscillations (shown by the blue star marker in the top panel of \hyperref[fig:fig3]{Fig. 3(a)}) for different $I$. We express the phase minima in terms of $\frac{\Phi_{\mathrm{ext}}}{\Phi_{\mathrm{period}}}$ and plot the phase minima points as a function of $I$ in \hyperref[fig:fig3]{Fig. 3(a)} (bottom panel), where $\Phi_{\mathrm{ext}} = B~A_{\mathrm{loop}}$ and $\Phi_{\mathrm{period}} = B_{\mathrm{period}}~A_{\mathrm{loop}}$. We now focus on the intercept of the extrapolated line and find that for the data shown in \hyperref[fig:fig3]{Fig. 3(a)} the intercept is approximately zero, modulo the period of oscillation ($\frac{\Phi_{\mathrm{ext}}}{\Phi_{\mathrm{period}}}$), corresponding to a phase difference $\phi_{12} \approx 0$ modulo $2\pi$. This implies that the two JJs in the SQUID are in a similar state. Now, we use another experimental knob available in our cryostat to apply a magnetic field in the plane of the device that is independently controlled. We sweep the in-plane magnetic field from $-20~\upmu\mathrm{T}$ to $20~\upmu\mathrm{T}$ and sweep it back, both sweeps in steps of $10~\upmu\mathrm{T}$. At each value of the in-plane field, we perform a similar flux modulation experiment with a perpendicular magnetic field. \hyperref[fig:fig3]{Fig. 3(b)} shows the result of such an experiment. Here, we also track the minima and extract an intercept. We find it to be close half-integer value of the oscillation period ($\frac{\Phi_{\mathrm{ext}}}{\Phi_{\mathrm{period}}}$). This corresponds to $\phi_{12} \approx\pi$, indicating that the two JJs are not in the same state -- this is a crucial observation of our SQUID-based experiment, and we will discuss the physical meaning next. \hyperref[fig:ext_fig2]{Extended Data Fig. 2} shows the result of measurements of $\phi_{12}$ as a function of inplane magnetic field. Data from another device is shown in \textbf{ SI 14.}

Now we try to understand our experimental observation of an anomalous phase difference $\phi_{12}$. The two Josephson junctions in the SQUID are realized by cutting the same twisted region; therefore, one would expect any phases arising from the symmetry of the order parameter across the two JJs to be the same. Hence, $\phi_{12}=0$ is expected and there is no net phase difference to be measured. However, crucially, the interfacial order parameter can be different in the two JJs. We understand this within the theoretical framework of Can \textit{et al.} \cite{can_high-temperature_2021}.  When one forms a twisted cuprate JJ, the emergent interfacial order is $\psi_1$+ $e^{i \alpha} \psi_2$ where $\psi_1$ and $\psi_2$ are the order parameters of the two BSCCO flakes; $\alpha$ is the quantum mechanical phase across the JJ. Using Landau-Ginzburg analysis of the free energy, Can \textit{et al.} \cite{can_high-temperature_2021} find that at a 45\textdegree~twist, the free energy has two degenerate minima, and by spontaneous symmetry breaking the system can be in one of these two minima. These two minima in free energy occur when the phase $\alpha$ is $\pi/2$ or $-\pi/2$. $\alpha=\pm\pi/2$ corresponds to an interfacial order of $d\pm id$ -- chiral superconductors of opposite chirality. This phase drop across the single JJ is not measurable. However, because we use a SQUID geometry, we can measure this phase difference when the two JJs in the loop have different phase $\alpha$  of $\pi/2$ and $-\pi/2$,  or $-\pi/2$ and $\pi/2$, across the two JJs, respectively, resulting in a net phase difference $\phi_{12}$ of $\pi$ or $-\pi$. We note that $\phi_{12}=\pi$ and $\phi_{12}=-\pi$ are equal to $\pi$  modulo 2$\pi$. Our theoretical calculations for the SQUID based on free energy analysis (see \textbf{SI 10}) are consistent with this expectation. 
Our experimental observations confirm this by measuring the phase difference in multiple devices as the in-plane magnetic field is varied. When the two junctions are in the same chiral state, then the relative phase difference is 0. The observation of different $\phi_{12}$ and associated orders is shown in \hyperref[fig:fig3]{Fig. 3(c and d)}. The ability to discern this phase difference $\phi_{12}$ comes from our unique SQUID geometry. The role of the in-plane magnetic field in controlling the distinct chirality of the two JJs is not well understood at this point. We propose the possibility that the in-plane field \cite{grigorenko_one-dimensional_2001,ghosh_high-temperature_2024} can result in flux focusing due to the coupling of Josephson and Abrikosov vortices and modify the free-energy landscape of the two JJs differently. 

\begin{figure*}[h]
\includegraphics[width=15.5cm]{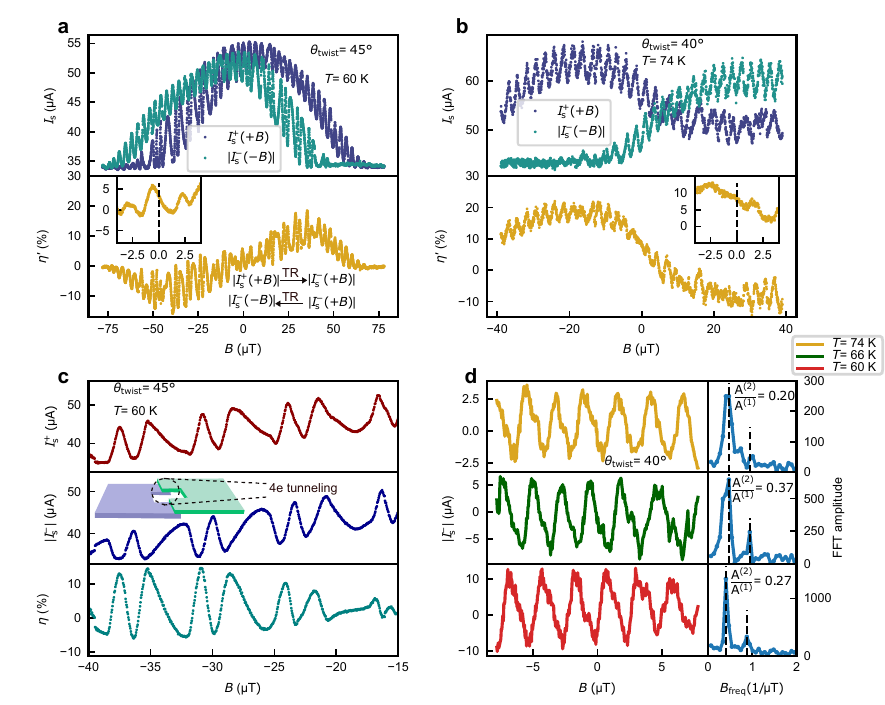}
\caption{ \label{fig:fig4} \textbf{Time-reversal symmetry breaking and signature of second harmonic content in the CPR.} (a) In the top panel, $I_s^+(+B)$ and $|I_s^-(-B)|$ are plotted together with the applied magnetic field for a 45\textdegree ~twisted device at 60 K. In the bottom panel, the corresponding field dependence of asymmetry $\eta'$ is plotted. $I_s^{+}(+B)$ and $|I_s^{-}(-B)|$ are connected by two time reversal (TR) operations, as shown in the plot. Asymmetry at $B=0$ suggests the intrinsic time-reversal symmetry breaking. The inset shows the field dependence of $\eta'$ close to $B=0$. At $B=0$, $\eta'$ is finite ($\sim 4 \%$, marked by black dashed line). (b) $I_s^+(+B)$ and $|I_s^-(-B)|$ are plotted together with the applied magnetic field for a 40\textdegree ~twisted device at 74 K in the top panel, and the bottom panel shows the magnetic field dependence of $\eta'$. Inset plot shows finite $\eta'$ $\sim 8 \%$ at $B=0$. (c) Signature of second harmonic in the switching current (both in $I^+_{\mathrm{s}}$ and $|I^-_{\mathrm{s}}|$), measured for 45\textdegree ~twisted device at 60 K. Schematic of two Cooper pairs (4$e$) tunneling across AJJ is shown in the inset. (d) Temperature dependence of the background-subtracted switching current of a 40\textdegree ~twisted device for three different temperatures (see legend) is shown, and corresponding FFTs are shown in the right-side panel. The ratio of the FFT amplitudes for the second and first harmonic components is indicated in the plot.}
\end{figure*}

Next, we examine additional aspects of our experiment that show evidence for chiral superconducting order at the twisted interface. \hyperref[fig:fig4]{Fig. 4(a)} shows the switching currents $I_s^+(+B)$ and $I_s^-(-B)$ for a 45\textdegree~twisted device. Where $I_s^+(+B)$ is the positive bias switching current for applied magnetic field value $B$, and $I_s^-(-B)$ is the negative bias switching current for the same magnetic field value but with a flipped sign. We define an asymmetry parameter $\eta' = \dfrac{I_s^+(+B)-|I_s^-(-B)|}{I_s^+(+B)+|I_s^-(-B)|} \times 100 \%$ to examine whether the time-reversal symmetry is preserved at $B=0$. It is clear from the top panel of \hyperref[fig:fig4]{Fig. 4(a)}, there is time reversal symmetry breaking at $B=0$ as the \textit{switching currents are not the same at the zero magnetic field}. This results in a non-zero $\eta'$ at $B=0$ as shown in the bottom panel of \hyperref[fig:fig4]{Fig. 4(a)} (behavior of $\eta '$ around $B=0$ is shown in the inset). Similar time reversal asymmetry breaking at $B=0$ is seen in a 40\textdegree~twisted device whose data is shown in \hyperref[fig:fig4]{Fig. 4(b)}. This kind of time-reversal symmetry breaking is expected in the scenario depicted in \hyperref[fig:fig3]{Fig. 3(c) and 3(d)}.

We now address the question of how the chiral superconducting order emerges. Close to a 45\textdegree ~twist angle, Cooper pair tunneling is suppressed, as evidenced by the reduced critical current density of the Josephson junctions \cite{ghosh_high-temperature_2024,zhao_time-reversal_2023,lee_encapsulating_2023}. It has been proposed that as the Cooper pair tunneling is reduced, the co-tunneling process could provide the Josephson coupling and result in an enhanced co-tunneling periodicity in the current phase relationship \cite{can_high-temperature_2021, yuan_inhomogeneity-induced_2023}. Signatures of higher-order harmonics in the CPR are seen in the data shown in the top panel of \hyperref[fig:fig3]{Fig. 3(a) and 3(b)}. We also observe signatures of higher harmonics in switching currents, as shown in \hyperref[fig:fig4]{Fig. 4(c)} for a 45\textdegree~twisted device. \hyperref[fig:fig4]{Fig. 4(d)} shows the evolution of the background-subtracted switching current as a function of the magnetic field at different temperatures for a 40\textdegree~device. The periodic oscillations on top of the smooth background indicate a distinct structure corresponding to the second-harmonic content in the CPR. To observe the temperature dependence of the first and second harmonic content, we have performed the FFT of the background-subtracted switching current modulation for the 40\textdegree~twisted device, shown in the right panel of \hyperref[fig:fig4]{Fig. 4(d)}.
The ratio of the FFT amplitude of the second harmonic ($A_2$) and the first harmonic ($A_1$) does not decrease monotonically (as shown in \hyperref[fig:fig4]{Fig. 4(d)} right panel) with temperature, which reflects the presence of simple Cooper pair transport along with the co-tunneling \cite{tummuru_josephson_2022}. The relative contribution of these two mechanisms determines the ratio of the FFT amplitude of two harmonics. Within the twist angle range of 40° to 45°, the second harmonic contribution increases as the angle approaches 45°, as shown in \textbf{SI 12}. In \textbf{SI 10}, we show a theoretical simulation of the CPR for different contributions of co-tunneling in the presence of finite geometric inductance.

We show unique evidence of emergence of chiral superconducting order at the twisted interface from the anomalous phase difference measured using the SQUID. Notably, this phase measurement is not possible using a single JJ \cite{zhao_time-reversal_2023,ghosh_high-temperature_2024,lee_encapsulating_2023, martini_twisted_2023} and is accessible only using SQUID. Additionally, two other aspects of our experiment -- the breaking of time-reversal symmetry at a zero magnetic field and the presence of higher harmonic content in switching currents -- are consistent with previous experiments \cite{zhao_time-reversal_2023}.
Recent theoretical work has suggested the possibility of inhomogeneity at the interface, resulting in breaking the time-reversal symmetry \cite{yuan_inhomogeneity-induced_2023}. In light of this, our experiment provides clarity by measuring direct evidence for chiral superconductivity from quantum interference measurements rather than relying solely on the presence of second harmonic in the CPR and diode response.

\begin{figure*}[h]
\includegraphics[width=15.5cm]{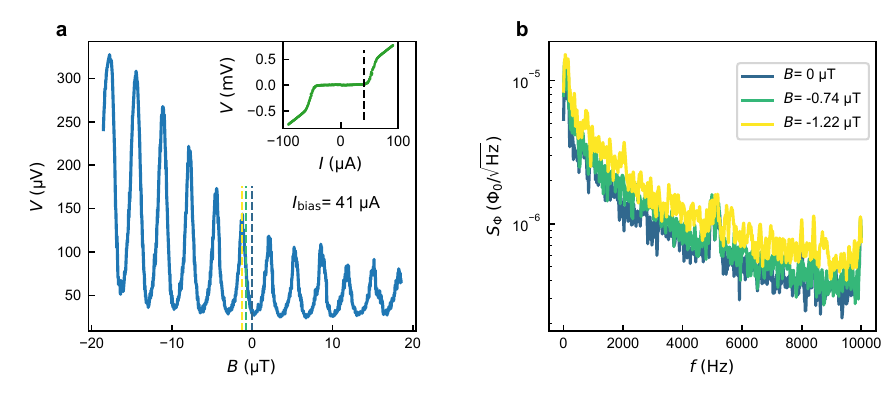}
\caption{ \label{fig:fig5} \textbf{Flux noise measurement of 45\textdegree~twisted BSCCO SQUID.} (a) Modulation of the SQUID voltage as a function of the applied perpendicular magnetic field. The inset shows the current biasing point near the critical transition in \textit{dc} $I-V$ curve. (b) The flux noise spectrum at distinct field biasing points, marked by the same color in (a). The flux noise is measured with a 50 Hz bandwidth.}
\end{figure*}

Although the physics of emergent interfacial superconducting order in these twisted SQUID devices is intriguing, these twist-angle-controlled SQUIDs also open up exciting avenues for future applications. In this work, we have demonstrated a first-of-its-kind c-axis SQUID. We measure the performance of these SQUIDs as a low-noise flux sensor at close to liquid $\mathrm{N_2}$ temperature. \hyperref[fig:fig5]{Fig. 5(a)} shows the transfer characteristics of the SQUID at 60 K, where the device is biased close to the superconducting to normal transition. SQUID performance can be characterized by parameters such as flux noise sensitivity $S_{\Phi}$, the most fundamental metric that depends on the underlying physics of the JJ. 
The frequency dependence of the flux noise sensitivity is shown in \hyperref[fig:fig5]{Fig. 5(b)} for three different field bias points, as shown by three dashed lines in \hyperref[fig:fig5]{Fig. 5(a)}. We find that the flux noise sensitivity of the SQUID is $\sim$1.5 $\mathrm{\upmu \Phi_0/\sqrt{Hz}}$ at 60 K. In \hyperref[fig:ext_fig1]{Extended Data Fig. 1}, we compare the flux‑noise and field‑noise sensitivity of our twisted BSCCO SQUID with those of previously reported low‑$T_c$ \cite{liu_temperature-dependent_2017,chen_high-performance_2016} and high‑$T_c$ SQUIDs \cite{dantsker_addendum_1995,faley_high-_2017}. From the comparison, we find that the flux‑noise sensitivity of our device is similar to that of state-of-the-art SQUIDs. A state-of-the-art flux sensor, such as the one demonstrated by us, can be converted to a highly sensitive magnetometer by further engineering of the loop area. Details of the flux noise measurement protocol are shown in \textbf{SI 5}.

In summary, the SQUID architecture is
 capable of measuring subtle quantum properties as it is sensitive to quantum interference. We observe an anomalous phase difference that can arise only when two distinct chiral superconducting orders are present in the two arms of the SQUID-- something not possible using single JJ experiments. Together with signatures of broken time-reversal symmetry at $B=0$ and higher-order components in the switching currents, we provide multiple complementary pieces of evidence for the chiral superconducting order. On a practical level, our SQUIDs have performance comparable to that of the state-of-the-art flux sensors. 

The signatures of interfacial superconductivity are subtle and can also lead to complex scenarios involving domains \cite{fermin_time-reversal_2025} that we cannot discern using SQUIDs in the present work. The analysis and device nanofabrication route we present would be naturally applicable to other van der Waals-based superconducting systems as well.

\clearpage

\section{Methods}

\textbf{Device Fabrication:} We fabricate twisted BSCCO SQUIDs employing the cryogenic exfoliation method \cite{zhao_time-reversal_2023} in an Ar-filled glovebox. First, we use scotch tape to exfoliate BSCCO flakes on a plasma cleaned Si/SiO2 substrate, which are heated overnight to remove any kind of adsorbed moisture on it. We select flakes with a thickness of 50-100 nm for the cryogenic exfoliation process. Inside the glovebox, we have built up a cryogen-compatible bottom stage, where we place our substrate using thermal grease, with exfoliated BSCCO flakes. PDMS stamps are employed for the cryogenic re-exfoliation process. We attach the stamp to a glass slide using Kapton tape and mount it onto the top stage dial. Next, we perform cryogenic re-exfoliation. In this step, the bottom stage is cooled by flowing liquid N2, while keeping the stamp in contact with the targeted flake. The PDMS stamp cleaves the BSCCO flake into two daughter flakes, near its glass transition temperature (-120°C). We then quickly rotated the top stage to the desired angle (with an accuracy of 0.5°) and dropped the top flake onto the bottom flake, creating the complete twisted stack. This entire stack fabrication process is performed very rapidly ($\sim$1 minute). Next, we make the Au contacts on the twisted stack using a SiN stencil by electron beam evaporation. This is followed by the deposition of Cr and Au through a metal mask to extend the electrodes. In the second evaporation step, we note that Cr is not deposited on the BSCCO stack. After fabricating the entire device, we carefully cut through the twisted overlap region with a sharp fiber scalpel \cite{varma_sangani_facile_2020}, giving it a SQUID geometry. We prepare these fiber scalpels using a Fujikura 80S fiber splicer. These tips are prepared using an electrical arc method and have a typical diameter of 1 $\upmu$m. Finally, we load the device immediately into the CRX-6.5 K closed-cycle cryostat for measurement. More details on the cryogenic re-exfoliation process are provided in \textbf{SI 1}, and we have provided a supplementary video of SQUID loop cutting.

\textbf{Measurement protocol:}
We have performed a device characterization using $R$ vs $T$ and dc $I-V$ measurements. To observe SQUID modulation, we performed dc-fixed current bias measurements and counter-based switching measurements. To detect the anomalous phase difference between two JJ arms of the SQUID, we performed $dV/dI$ measurements with a perpendicular magnetic field for different fixed \textit{dc} current biases. SQUID noise measurements are done using a spectrum analyzer. Here, we discuss the method of basic device characterization. Details about the other measurement protocol are provided in \textbf{SI 5}.

\textbf{$R$ vs $T$ and \textit{dc} $I-V$ measurements:}
$R$ vs. $T$ characteristics is the initial test of the device that confirms the superconductivity of the junction. We apply a \textit{ac} current of magnitude 100 nA by using the SRS 830 lock-in and a 1 M$\Omega$ ballast resistor. Simultaneously, we measure the voltage drop across the junction and across the pristine flake using SRS 830 lock-in after amplification using a DL Instruments 1201 preamplifier. We measure this voltage drop while ramping down the cryostat temperature using the Lake Shore 336 temperature controller.
We use the NI Data Acquisition System (DAQ) for \textit{dc} \textit{I-V} measurements. We use the analog output channel of the DAQ and a 10 k$\Omega$ ballast resistor to send a \textit{dc} current through the device. We also measure the current using a current-to-voltage converter and the DAQ. Simultaneously, we measure the voltage drop across the junction using DAQ after amplifying it with the DL Instruments pre-amplifier. 

\textbf{Connection between two phase sum rule equation:}
We have used the two-phase sum rule equations \hyperref[eq:eq1]{Eq. 1} and \hyperref[eq:eq2]{Eq. 2} to understand the SQUID response in two different contexts. In \hyperref[eq:eq1]{Eq. 1}, we did not include the anomalous phase difference between the two JJ arms, since the goal of that equation is to extract the Josephson inductance from the slope, and a constant phase difference does not affect the slope. 
In \hyperref[eq:eq2]{Eq. 2} we have replaced the term $2\pi(\dfrac{(L_{geo,1}I_{1}(\phi_1) - L_{geo,2}I_{2}(\phi_2))}{\Phi_0})$ with $\mathcal{A}~L_{geo}~I$ and included the anomalous phase difference $\phi_{12}$ between two JJ arms. Below, we show how these two terms are connected to each other.

Defining the geometric inductance asymmetry $\nu$ and current asymmetry $\gamma$ between two JJ arms of the SQUID one obtains: $L_{geo,1}= \dfrac{L_{geo}}{2}(1+\nu)$, $L_{geo,2}= \dfrac{L_{geo}}{2}(1-\nu)$ and $I_{1}= \dfrac{I}{2}(1+\gamma)$, $I_{2}= \dfrac{I}{2}(1-\gamma)$. Putting these values in \hyperref[eq:eq1]{Eq. 1}, the expression $2\pi(\dfrac{(L_{geo,1}I_{s1}(\phi_1) - L_{geo,2}I_{s2}(\phi_2))}{\Phi_0})$ becomes $\dfrac{L_{geo} I}{2}(\nu+\gamma)$. From the comparison with \hyperref[eq:eq2]{Eq. 2}, term $\mathcal{A}$ turns out $\dfrac{2\pi}{\Phi_0}(\dfrac{\nu+\gamma}{2})$. $\mathcal{A}~L_{geo}~I$ term can be written in terms of screening parameter $\beta$, as $\beta \dfrac{I}{2I_c}(\nu+\gamma)$. Physically $\nu$ suggests the perimeter asymmetry and $\gamma$ suggests the area asymmetry between the two arms. As in the simplest approximation, the geometric inductance depends on the perimeter of the arms, and the current depends on the area of the arms so approximately we can write $\nu \approx \frac{P_1-P_2}{P_1+P_2}$ and $\gamma \approx \frac{A_1-A_2}{A_1+A_2}$, where $P_{1(2)}$ and $A_{1(2)}$ is the perimeter and the area of the JJ arm 1(2).  

\section*{Acknowledgments}

We thank Eli Zeldov, Philip Kim, Vibhor Singh, Sophie Gueron, Hélène Bouchiat, Shamashis Sengupta, R. Vijay, Arti Garg, Felix von Oppen, and Vladimir Krasnov for helpful discussions and comments. M.M.D. acknowledges the Department of Science and Technology (DST) of India for J.C. Bose fellowship JCB/2022/000045. M.M.D acknowledges funding support for SPIKE from the National Mission on Interdisciplinary Cyber Physical Systems, of the Department of Science and Technology, Govt. of India through the I-HUB Quantum Technology Foundation I-HUB/SPIKE/2023-24/009. We acknowledge support from CEFIPRA CSRP Project no. 70T07-1. We acknowledge support from AOARD Project no. FA2386-25-1-4027. Department of Atomic Energy (DAE) of Government of India 12-R\&D-TFR-5.10-0100 for support.

\section*{Author contributions}
A.B., S.A., and N.B. fabricated the devices. S.N., T.G., A.D. and K.A. helped in device fabrication. A.B. did the measurements. A.S. helped in the SiN membrane making. D.A.J., Ruta K., and A.T. grew the BSCCO crystals. J.S. helped in the AWR microwave simulation and noise measurements. A.B. and M.M.D. did the numerical analysis. S.A., S.N., and T.G. helped in the analysis. Ritajit K. and A.K. did the theoretical calculations. A.B. and M.M.D. wrote the manuscript. All authors provided inputs to the manuscript. M.M.D. supervised the project.
\section*{Competing interests}
The authors declare no competing financial interests.

\newpage

\renewcommand{\figurename}{\textbf{Extended Data Fig.}}   
\renewcommand{\thefigure}{\number\numexpr\value{figure}-5\relax}

\begin{figure}[htbp]
    \centering
    \caption{\textbf{Flux and field noise sensitivity comparison of high $T_c$ and low $T_c$ SQUIDs with our work.}}
    \label{fig:ext_fig1}
    \begin{tabular}{|c|c|c|c|c|c|c|}
        \hline
        Report & Material & Flux noise & Field noise & Pick up coil & Area & Operating \\
        & & ($\mu \Phi_0$/$\sqrt{\mathrm{Hz}}$)&& & SQUID loop/coil & temp (K) \\
        \hline
        Liu et al. \cite{liu_temperature-dependent_2017} &
        NbN &
        3.9 &
        6.6 fT/$\sqrt{\mathrm{Hz}}$ &
        Pickup coil &
        29.12 mm$^2$&
        9 K \\
        \hline
        Chen et al. \cite{chen_high-performance_2016} &
        Nb &
        0.34 &
        75 pT/$\sqrt{\mathrm{Hz}}$ &
        No &
        9 $\mu$m$^2$&
        4.5 K \\
        \hline
        Faley et al. \cite{faley_high-_2017} &
        YBCO &
        10 &
        5 fT/$\sqrt{\mathrm{Hz}}$ &
        Flip-chip &
        4.6 mm$^2$&
        77.5 K \\
        \hline
        Dantsker et al. \cite{dantsker_addendum_1995} &
        YBCO &
        4.9 &
        8.5 fT/$\sqrt{\mathrm{Hz}}$ &
        Flip chip &
        1.2 mm$^2$&
        77 K \\
        \hline
        \textcolor{red}{Our work} &
         Twisted BSCCO &
        1.41 &
        34 pT/$\sqrt{\mathrm{Hz}}$ &
        No &
        81 $\mu$m$^2$&
        77 K \\
        \hline
    \end{tabular}
\end{figure}

\begin{figure*}[h]
\includegraphics[width=15.5cm]{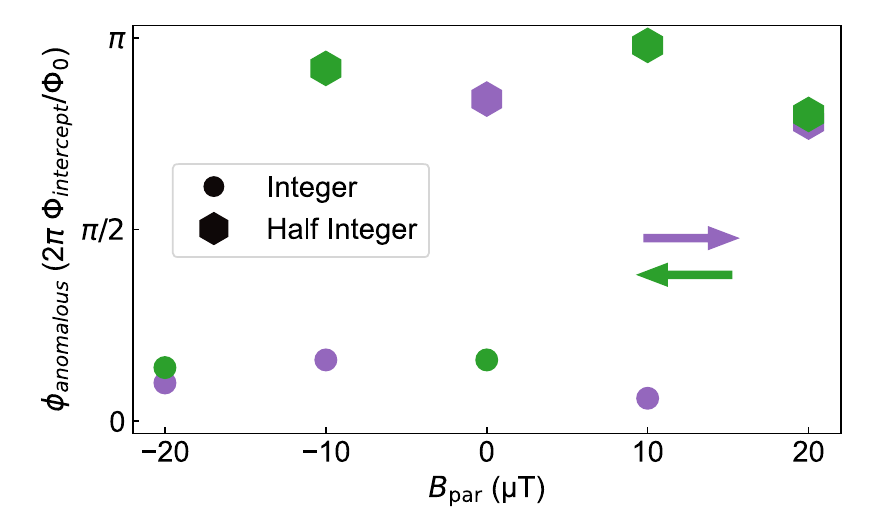}
\caption{ \label{fig:ext_fig2} \textbf{Anomalous phase statistics with in-plane magnetic field for 45\textdegree~twisted SQUID at 20 K.} Variation of modulo 2$\pi$ anomalous phase difference between two JJ arms with in-plane magnetic field is shown. Arrowhead shows the direction of the in-plane field sweep.}
\end{figure*}

\end{document}